\documentclass[conference]{IEEEtran}
\IEEEoverridecommandlockouts
\usepackage[english]{babel}
\usepackage{cite}
\usepackage[hidelinks]{hyperref} % Hyperlinks
\usepackage{amsmath,amssymb,amsfonts}
\usepackage{siunitx}
\usepackage{algorithmic}
\usepackage{graphicx}
\usepackage{textcomp}
\usepackage{xcolor}
\usepackage{siunitx}
\usepackage{booktabs}
\def\BibTeX{{\rm B\kern-.05em{\sc i\kern-.025em b}\kern-.08em
    T\kern-.1667em\lower.7ex\hbox{E}\kern-.125emX}}
\begin{document}

\title{Evaluating the Impact of Model Accuracy for Optimizing Battery Energy Storage Systems
    \thanks{
        This work was funded by the Bavarian Research Foundation under the research project
        KI-M-Bat (reference number AZ-1563-22).
    }
}

\author{
    Martin Cornejo\textsuperscript{1},
    Melina Graner\textsuperscript{1,2},
    Holger Hesse\textsuperscript{2},
    Andreas Jossen\textsuperscript{1} \\[0.1cm]
    \textsuperscript{1}Technical University of Munich, Germany;
    TUM School of Engineering and Design, \\
    Department of Energy and Process Engineering,
    Chair of Electrical Energy Storage Technology \\[0.1cm]
    \textsuperscript{2}Kempten University of Applied Sciences, Germany; Institute for Energy and Propulsion Technology
    \\[-0.5cm]
}
\maketitle

\begin{abstract}
    This study investigates two models of varying complexity for optimizing intraday arbitrage energy trading
    of a battery energy storage system using a model predictive control approach.
    Scenarios reflecting different stages of the system's lifetime are analyzed.
    The findings demonstrate that the equivalent-circuit-model-based non-linear optimization
    model outperforms the simpler linear model
    by delivering more accurate predictions of energy losses and system capabilities. This enhanced accuracy enables
    improved operational strategies, resulting in increased roundtrip efficiency and revenue,
    particularly in systems with batteries exhibiting high internal resistance, such as second-life batteries.
    However, to fully leverage the model's benefits, it is essential to identify the correct parameters.
\end{abstract}


\section{Introduction}
% motivation on BESS, flexibility and optimization
Battery energy storage systems (BESS) are essential for providing the flexibility needed
to continue expanding renewable power generation while maintaining grid stability.
The flexible nature of BESS also creates a need for
effective energy management to fully leverage the system's capabilities.
While several heuristic and machine learning methods have been explored as algorithmic foundations
for developing operational strategies,
mathematical optimization remains the most commonly used approach in this field \cite{weitzel_energy_2018}.
BESS optimization models aim to determine the best power schedule to maximize an objective % (typically economic)
while considering technical constraints, like power and state-of-charge (SOC) limits.
This approach provides a model-based framework that is physically interpretable
and easily transferable to new systems and applications.

% literature review - LP models
Similar to battery simulation models, battery optimization models range from simplified representations to complex
physicochemical models, each designed for specific applications and use cases \cite{rosewater_battery_2019}.
However, the vast majority of optimization models in the literature use linear programming (LP) formulations
\cite{weitzel_energy_2018}.
LP models take a simplified approach and construct an optimization model with constraints and objectives represented with linear relationships.
As a result, the optimization process is very fast,
making LP models particularly useful for applications that require high scalability and extended optimization horizons,
such as optimal system sizing \cite{hesse_economic_2017,martins_optimal_2018} and
transmission expansion planning \cite{pozo_linear_2022}.

% drawback LP models
While LP models provide globally optimal solutions within their feasible region,
their simplified assumptions limit their ability to capture real-world behavior,
potentially leading to suboptimal or infeasible schedules in actual systems.
Notably, LP models typically represent losses using a constant efficiency, whereas,
in reality, system losses typically grow quadratically with output power
and depend on operational conditions such as SOC and battery temperature.
Previous studies have focused on developing optimization models that more accurately reflect
this behavior, for example through convex relaxations \cite{despeghel_convex_2024}
and piecewise linear approximations \cite{hesse_ageing_2019, kumtepeli_energy_2020}.

% NL ECM models
Additionally, most BESS optimization models focus solely on power and energy,
while batteries are more accurately described with equations governing voltage, current and charge.
Previous studies have shown that using an equivalent circuit model (ECM) for optimization results in a
formulation with non-linear (NL) equations, which is more computationally intensive to solve.
However, compared to LP models, this approach can better estimate system energy losses
and help avoid schedules that lead to technical violations in the actual system
\cite{aaslid_non-linear_2020, rosewater_risk-averse_2020}.

% paper objectives
In this context, this paper aims to evaluate the advantages of using a high-fidelity
non-linear (NL) model over a simplified linear programming (LP) model
for optimizing BESS operations.
Additionally, it seeks to quantify the performance loss that occurs due to
discrepancies between model parameters and the actual system.
To illustrate these points, a case study is presented that focuses on the closed-loop optimization
of a BESS to maximize revenues in intraday energy trading.

\section{Optimization models}

\subsection{Linear model}
The main decision variables of the linear model
are system power $p_t$ and SOC $soc_t$ for each timestep in the optimization horizon $t \in T$.
The SOC updates based on the system power, the timestep length $\Delta t$,
the system's nominal energy capacity $E_N$,
and considers charge and discharge losses modeled from a constant efficiency factor $\eta$.
\begin{equation}
    soc_t = soc_{t-1} + \frac{\Delta t}{E_N} \, \left(p^{ch}_t \, \eta - p^{dch}_t \, \frac{1}{\eta}\right)
    \quad \forall t \in T \label{eq:lp-soc}
\end{equation}

The power is a combination from charging power $p^{ch}_t$ and discharging power $p^{dch}_t$ variables.
Although the model does not enforce mutual exclusivity between charge and discharge power,
it discourages simultaneous charging and discharging
as the additional energy losses would result in a suboptimal solution.
\begin{equation}
    p_t = p^{ch}_t - p^{dch}_t \qquad \forall t \in T \label{eq:lp-power}
\end{equation}
The power variables are limited by the system's rated power,
while the SOC is also constrained to defined lower and upper limits.
% safety margins
\begin{align}
    0         & \leq p^{ch}_t  \leq p^{max}   & \forall t \in T                         \\
    0         & \leq p^{dch}_t \leq p^{max}   & \forall t \in T \label{eq:lp-power-lim} \\
    soc^{min} & \leq soc_t     \leq soc^{max} & \forall t \in T \label{eq:lp-soc-lim}
\end{align}

\subsection{Non-linear model}
The non-linear model employs a simple ECM to model the battery.
The SOC is charge based instead of energy based, it is updated from current $i$ and
nominal charge capacity $Q_N$.
% Assuming a perfect coloumb efficiency.
\begin{equation}
    soc_t = soc_{t-1} + \frac{\Delta t}{Q_N} \, i_t \qquad \forall t \in T \label{eq:ecm-soc}
\end{equation}

The battery power $p^{dc}_t$ is the product from terminal voltage and current.
The terminal voltage emerges from
the battery's open-circuit voltage (OCV) and the voltage drop caused by the battery's internal resistance $R$.
The OCV is a function of SOC
and can be modeled, for example, through a polynomial curve or cubic splines \cite{rosewater_battery_2019}.
\begin{align}
    p^{dc}_t & = v_t \; i_t       & \forall t \in T  \label{eq:ecm-power}  \\
    v_t      & = ocv_t + i_t \; R & \forall t \in T \label{eq:ecm-voltage}
\end{align}

Equations (\ref{eq:ecm-power}) and (\ref{eq:ecm-voltage}) inherently account
for quadratically increasing losses with battery output power, which arise from ohmic losses $i^2 \, R$.
Additionally, the SOC also influences these losses,
as a higher SOC corresponds to a higher OCV, which in turn results in a higher efficiency
since less current is required to provide the same amount of power.

The battery power is constrained by its voltage and current limits, and the SOC also has defined boundaries.
\begin{align}
    -i^{max}  & \leq i_t \leq i^{max}     & \forall t \in T \\
    v^{min}   & \leq v_t \leq v^{max}     & \forall t \in T \\
    soc^{min} & \leq soc_t \leq soc^{max} & \forall t \in T
\end{align}

To achieve the desired voltage and energy, cells are connected in $s$ series and $p$ parallel configurations.
Consequently, the battery parameters, typically available at the cell level, must be scaled accordingly.
\begin{align}
    Q_N & = Q_{cell} \cdot p     &
    i   & = i_{cell} \cdot p       \\
    v   & = v_{cell} \cdot s     &
    R   & = R_{cell} \cdot s / p
\end{align}

In addition to the battery, the converter losses are also incorporated.
A simplified approach is to model them through a constant efficiency $\eta^{conv}$.
The converter power is constrained to its rated power as in equations
(\ref{eq:lp-power})-(\ref{eq:lp-power-lim}).
\begin{equation}
    p^{dc}_t  = p^{ch}_t \, \eta^{conv} - p^{dch}_t \, \frac{1}{\eta^{conv}} \qquad \forall t \in T \label{eq:ecm-conv}
\end{equation}

% Since the converter efficiency is typically also dependent on its operating conditions
% other possibilities are to model the losses through a polynomial or cubic splines
% \cite{rosewater_battery_2019,aaslid_non-linear_2020}.

\section{Case study}
\subsection{System and scenarios description}
\begin{figure*}
    \includegraphics[width=\linewidth]{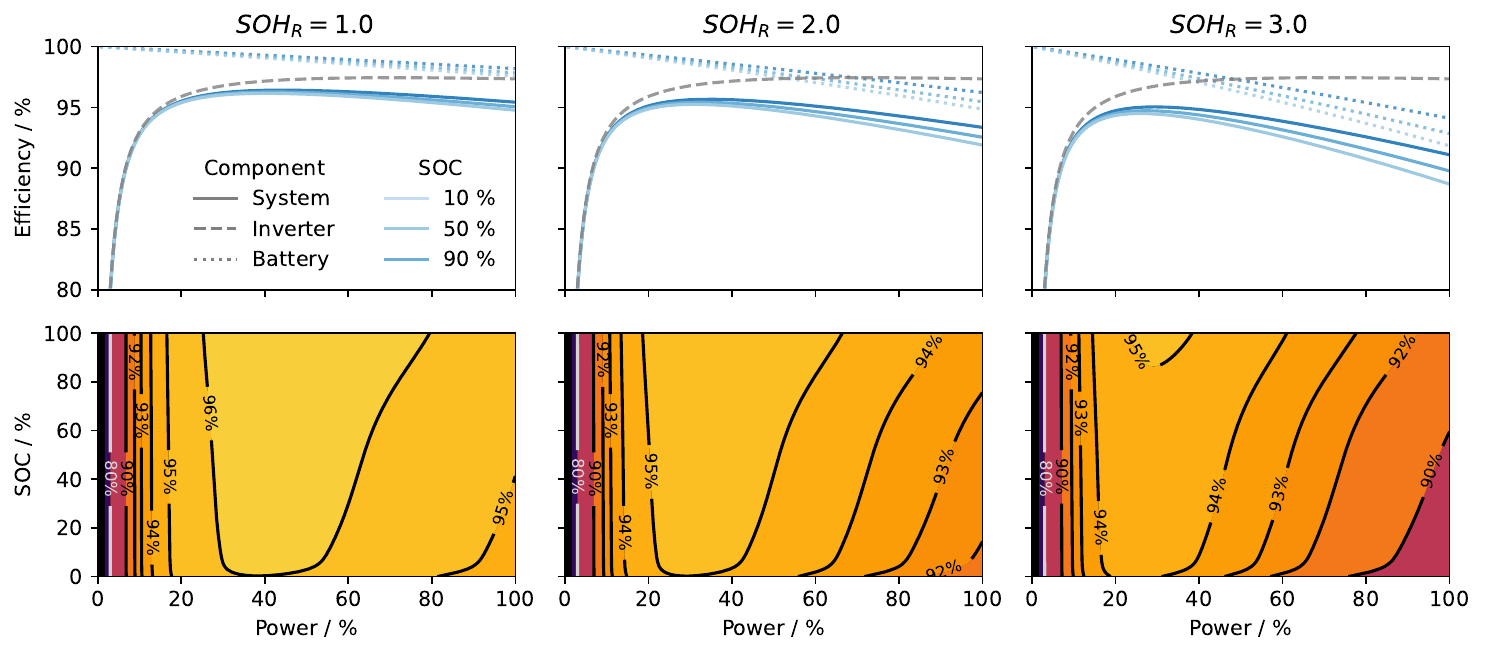}
    \caption{
        System efficiency as a function of output power (discharge direction)
        and SOC under increasing battery resistance conditions.
        The top section displays efficiency curves at the component level
        -- converter, battery, and overall system -- at SOC values of \SI{10}{\%}, \SI{50}{\%}, and \SI{90}{\%}.
        The bottom section presents a 2D contour map of the system efficiency
        across all operating conditions.
    }
    \label{fig:eff}
\end{figure*}

% system characteristics
The storage system considered this study has a power rating of \SI{180}{kW} and an energy capacity of \SI{180}{kWh}.
Multiple units of this size are typically aggregated to form larger, MW-scale energy storage systems.
Two main components comprise the system model: the converter, a Siemens Sinamics S120 characterized by
Schimpe et al. \cite{schimpe_energy_2018},
and the battery, made of Samsung SDI NMC cells characterized by Collath et al. \cite{collath_suitability_2024}.
Table \ref{tab:specs} provides an overview of the system characteristics.
\begin{table}[h]
    \centering
    \caption{System and Cell Specifications}
    \begin{tabular}{ll}
        \toprule
        % \textbf{Parameter}          & \textbf{Value}                        \\
        % \midrule
        \multicolumn{2}{c}{\textbf{System Specifications}}         \\
        \midrule
        Rated power                  & \SI{180}{kW}                \\
        Nominal energy capacity      & \SI{180}{kWh}               \\
        Nominal voltage              & \SI{972}{V}                 \\
        Usable SOC window            & \SI{0}{\%} - \SI{100}{\%}   \\
        Inverter model               & Sinamics S120               \\
        Cell type                    & NCM Samsung SDI \SI{94}{Ah} \\
        Cell configuration           & 260s2p                      \\
        \midrule
        \multicolumn{2}{c}{\textbf{Cell Specifications}}           \\
        \midrule
        Nominal capacity             & \SI{94}{Ah}                 \\
        Nominal voltage              & \SI{3.68}{V}                \\
        Voltage limits               & \SI{2.7}{V} - \SI{4.15}{V}  \\
        Internal resistance          & \SI{0.819}{\milli\ohm}      \\
        Max. charge/discharge rating & 2C / 2C                     \\
        \bottomrule
    \end{tabular}
    \label{tab:specs}
\end{table}

% scenarios
Three scenarios are evaluated, each representing a different stage in the system's lifetime
and the consequent impact on the system's efficiency.
To this end, the cell's internal resistance is varied based on its state of health, defined as follows:
\begin{equation}
    SOH_R = \frac{R}{R_{BOL}}
\end{equation}
where $R_{BOL}$ is the internal resistance at beginning-of-life (BOL).
The cell is a large-format type intended for stationary systems and exhibits an internal
resistance of \SI{0.819}{m\Omega} at BOL.
This resistance was characterized at \SI{25}{\celsius} and represents
an average value across the measured SOC range \cite{collath_suitability_2024}.
% the OCV curve-fit from measurements at 25°C

The first scenario, with $SOH_R=1.0$, considers a system at BOL.
The second scenario, with $SOH_R=2.0$, represents highly aged cells after about a decade in operation.
The third scenario, with $SOH_R=3.0$, examines the extreme case with a second-life battery system.

% efficiency characterization
For all three scenarios the system efficiency as a function of output power and SOC was characterized, with
the results presented in Figure \ref{fig:eff}.
The battery losses increase quadratically, leading to a linear decreasing efficiency.
Lower SOC levels contribute to this decline in efficiency.
The real converter exhibits a poor efficiency at low partial loads
but maintains a higher and almost constant efficiency at higher power levels.
The combined efficiency curve, shows lower efficiency at both partial loads and high power ratings.
With increasing cell resistance, the battery losses increase, resulting a more pronounced efficiency curve.

\subsection{Benchmark}
The two optimization models are compared across the three scenarios
in the context of arbitrage trading in the German continuous intraday electricity spot market.
The strategy involves generating profit by buying (charging) at low electricity prices $c_t$
and selling (discharging) later at a higher price.
\begin{equation}
    \min \sum_{t \in T} c_t \,  p_t \, \Delta t
\end{equation}
Each model's objective is to maximize revenues, subject to the technical constraints of the system, which are
represented by the constraints of the LP model (\ref{eq:lp-soc})-(\ref{eq:lp-soc-lim})
and the NL model (\ref{eq:ecm-soc})-(\ref{eq:ecm-conv}), respectively.
To avoid excessively fast aging of the batteries,
the system throughput is limited to $1.5$ full equivalent cycles (FEC) per day,
a common practice that reflects typical warranty requirements.

To replicate real conditions, the optimization models are implemented and evaluated
within a model predictive control (MPC) framework.
For this a BESS simulation model is introduced to emulate the
response of a real system to the optimizer signals.
The process works as follows: the optimizer creates a power dispatch schedule for the upcoming optimization horizon,
the \textit{real system} executes part of this plan, before the optimizer
reschedules again based on updated price forecasts and the current SOC as initial condition.

The performance of the optimizers is benchmarked through a one-year simulation.
A perfect price forecast is assumed, utilizing the historical ID-1 prices of 2021, which has been identified as
a suitable index for benchmarking BESS trading \cite{hornek_value_2025}.
The optimization horizon is \SI{12}{\hour}, with an action horizon of \SI{15}{\minute}.
Both the optimization and simulation are modelled with a \SI{1}{\minute} timestep discretization.

All model parameters are derived from the system specifications, except
the system efficiency $\eta$  for the LP model and the converter efficiency $\eta^{conv}$ for the NL model.
The parameters are identified by fitting them to the energy losses along the full efficiency curve at \SI{50}{\%} SOC
for each $SOH_R$ scenario (Fig. \ref{fig:eff}),
resulting in efficiencies of \SI{95.9}{\%}, \SI{94.6}{\%} and \SI{93.3}{\%}.
The converter efficiency $\eta^{conv}$ is fitted similarly, excluding the battery losses,
yielding a value of \SI{97.3}{\%}.

The performance metrics include revenue, roundtrip efficiency and energy shortfall.
Roundtrip efficiency is the ratio of the total energy output during
discharge phases to the total energy input during the charging phases, taking into account the initial and final SOC.
\begin{equation}
    \eta_{RTE} = \frac{E_{out}}{E_{in} - E_N (\text{SOC}_{\text{end}} - \text{SOC}_{\text{start}})}
\end{equation}
The energy shortfall measures the total discrepancy between scheduled power $p^{opt}_t$
and actual delivered power $p^{real}_t$.
\begin{equation}
    E_{imb} = \sum_t | p^{opt}_t - p^{real}_t| \Delta t
\end{equation}

In addition, a sensitivity analysis is conducted to evaluate the change in revenue and energy shortfall
for each scenario as a result of varying optimization model parameters.
Since these parameters are not always readily available or easily identifiable,
the goal of the analysis is to determine how discrepancies between the optimization model parameters
and the actual system parameters affect overall performance.
For the LP model, the efficiency $\eta$ is varied between \SI{90}{\%} and \SI{97}{\%}.
For the NL optimization model, the internal resistance $R$ is underestimated and overestimated
by varying it with a factor ranging from \SI{50}{\%} to \SI{150}{\%}.

The optimization models are implemented in Pyomo
and optimized using the HiGHS solver for the linear model and the Bonmin solver for the non-linear model.
The system response is simulated with SimSES \cite{moller_simses_2022}
using the battery ECM and full converter efficiency curve.
The code implementation is available as open-source \cite{github}.

\section{Results}
\subsection{Model comparison}
\begin{figure}
    \includegraphics[width=\linewidth]{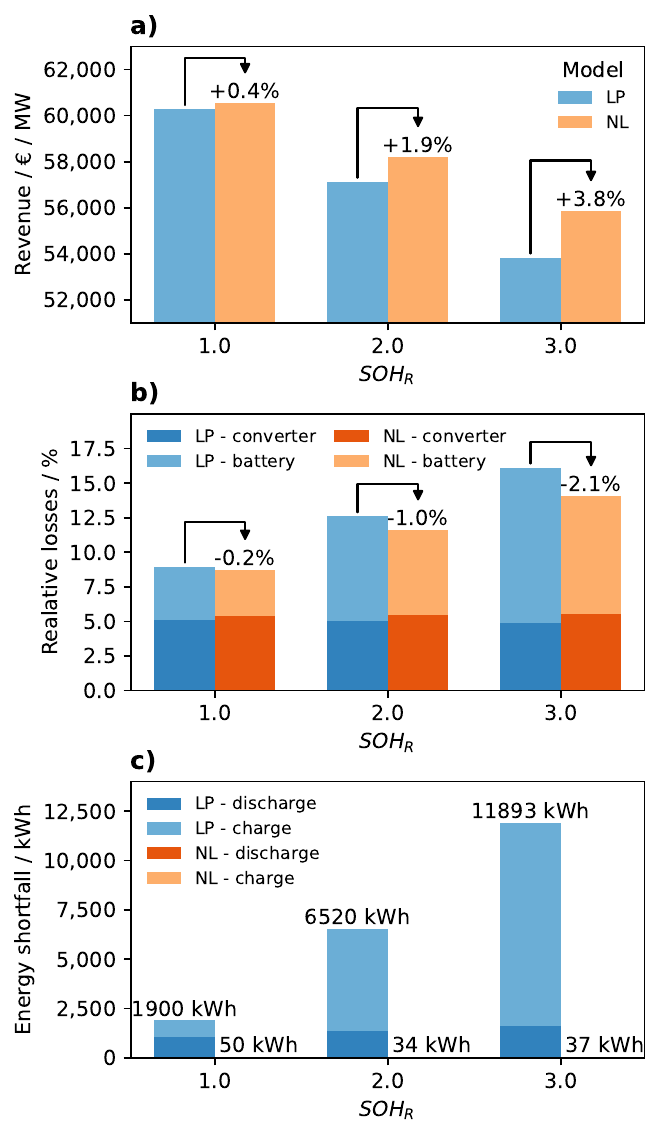}
    \caption{
        Benchmark results for LP model and NL model under scenarios with increasing battery internal resistance.
        \textbf{a)} Yearly revenue in Euros per MW.
        \textbf{b)} Relative energy losses from battery and converter.
        \textbf{c)} Energy shortfall due to unfulfilled power target setpoints.
    }
    \label{fig:benchmark}
\end{figure}

% system revenue and roundtrip efficiency
The results are summarized in Table \ref{tab:benchmark} and Figure \ref{fig:benchmark}.
In terms of revenue, both models demonstrate comparable performance in the $SOH_R=1.0$ scenario,
with around \SI{60000}{\text{€}\per\mega\watt} generated yearly.
As internal resistance increases, revenue is impacted.
However, the LP model experiences a more rapid decline and by the $SOH_R=3.0$ scenario the NL model shows a
significant advantage.
The same behavior is observed for the roundtrip efficiency, indicating a strong relationship.
Specifically, a \SI{1}{\%} loss in RTE is associated with an approximate \SI{1.5}{\%} loss in revenue,
supported by a correlation coefficient of $r=0.998$.

\begin{table}[b]
    \setlength{\abovecaptionskip}{1pt}
    \centering
    \caption{Benchmark results}
    \begin{tabular}{ccrrr}
        \toprule
        \multicolumn{2}{c}{\textbf{Scenario}}
                         & \textbf{Revenue} & \textbf{Roundtrip} & \textbf{Energy imb.}                   \\
        \textbf{$SOH_R$} & \textbf{Model}   & \textbf{in €/MW}   & \textbf{efficiency}  & \textbf{in kWh} \\
        \midrule
        $1.0$            & LP               & $60,278$           & $\SI{91.4}{\%}$      & $1,900$         \\
                         & NL               & $60,544$           & $\SI{91.5}{\%}$      & $50$            \\
        $2.0$            & LP               & $57,122$           & $\SI{87.7}{\%}$      & $6,520$         \\
                         & NL               & $58,211$           & $\SI{88.6}{\%}$      & $34$            \\
        $3.0$            & LP               & $53,833$           & $\SI{84.2}{\%}$      & $11,893$        \\
                         & NL               & $55,867$           & $\SI{86.1}{\%}$      & $37$            \\
        \bottomrule
    \end{tabular}
    \label{tab:benchmark}
\end{table}

% relative losses / system efficiency
Across all scenarios, the relative converter losses remain nearly constant,
while battery losses grow proportionally with their internal resistance increase (Fig. \ref{fig:benchmark}b).
This results in the system efficiency curve deviating further from a constant efficiency,
making the LP model assumptions increasingly inaccurate.
In contrast, the NL model is adaptive to the cell's $SOH_R$ and adjusts its strategy accordingly.

\begin{figure}
    \includegraphics[width=\linewidth]{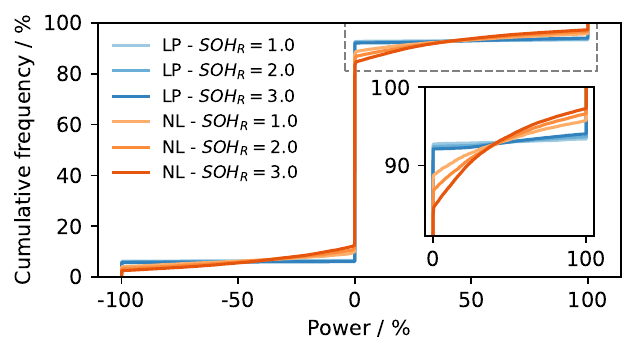}
    \caption{Cumulative distribution function of BESS power for
        the LP model and NL model under scenarios with increasing battery internal resistance.
    }
    \label{fig:power}
\end{figure}

% power distribution
This is evident from the power distribution from the BESS simulations, as shown in Fig \ref{fig:power}.
The LP model strategy remains invariant of scenario, charging and discharging at full power whenever possible.
% The efficiency parameter only informs the model with the marginal costs associated with the energy losses,
% which has only a minimal impact between scenarios.
In contrast, the NL model, aware of the high losses associated with operating at full power,
distributes the output more evenly across the power range.
As resistance increases, the NL model increasingly aims to minimize high power levels
and allocates more power to lower levels.


% energy shortfall / forecasted revenue vs. real revenue
The NL model also demonstrates significantly better predictive performance regarding system capabilities,
showing minimal discrepancies between the scheduled and actual power dispatch across all scenarios.
In contrast, the LP model experiences substantial power deviations that increase with rising battery resistance.
For instance, in the scenario with $SOH_R = 3.0$ the energy shortfall is equivalent
to around \SI{5}{\%} of the total energy throughput.
% The charge direction is particularly affected.
% As resistance increases, voltage limitations are reached at lower SOC levels,
% thereby restricting power delivery capabilities that the LP model does not account for.
These unexpected shortfalls can lead to further economic repercussions in energy trading,
as markets often penalize disparity between scheduled and delivered power.
Even more, accurately predicting system capabilities is critical for applications where not delivering the
required power can have severe consequences, such as in the provision of grid ancillary services.

\subsection{Sensitivity analysis}
\begin{figure}
    \includegraphics[width=\linewidth]{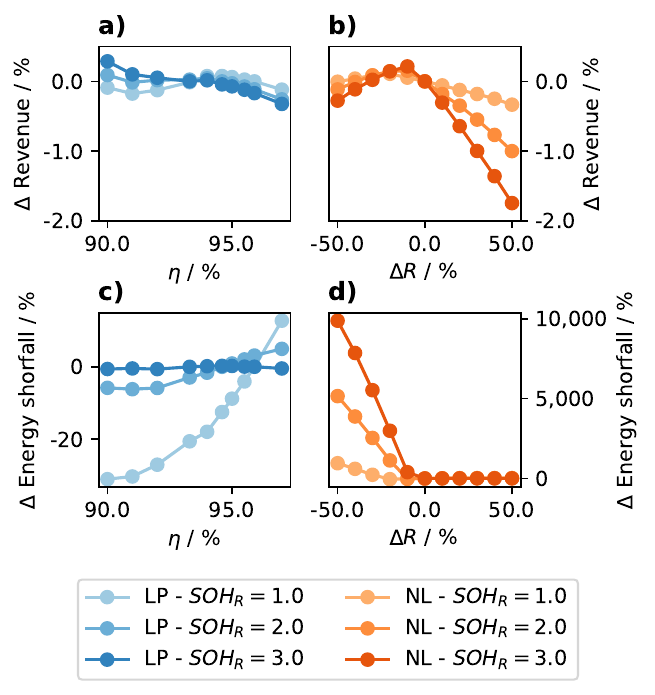}
    \caption{Sensitivity analysis:
        revenue loss and increase in unfulfilled power schedules compared to baseline scenarios,
        resulting from varying efficiency values $\eta$ for the LP optimization model
        and discrepancies between the battery resistance $R$ of the NL optimization model and the real system.
    }
    \label{fig:sensitivity}
\end{figure}
% Figure \ref{fig:sensitivity} presents the results of the sensitivity analysis.
Varying the efficiency parameter $\eta$ has a limited effect on the LP model's performance.
Pessimistic values help to counterbalance the model's overly optimistic nature.
Consequently, the energy shortfall can be slightly reduced, leading to marginal improvements in revenue
(Fig. \ref{fig:sensitivity}a and \ref{fig:sensitivity}c).
This reveals a contradiction in selecting $\eta$,
as the physically meaningful efficiency does not always correspond to the optimal results.

The NL model is much more sensitive to deviations from the true internal resistance $R$.
Underestimating the internal resistance leads to an overly optimistic prediction
of the energy losses and system capabilities, significantly increasing the energy shortfall
(although still remaining below that of the LP model) and negatively affecting revenues.
In the same way, overestimating the internal resistance results in overly pessimistic estimation.
While the energy shortfall is not affected, the system does not make use of its full capabilities,
causing the advantage in the revenues diminish rapidly
as the model parameters deviate further from the \textit{true} values
(Fig. \ref{fig:sensitivity}b and \ref{fig:sensitivity}d).
This highlights the importance of accurately determining the correct parameters to ensure
the model's reliability and optimal performance.

\section{Conclusion and Outlook}
This study evaluated two models of varying complexity for optimizing intraday arbitrage energy trading
of a BESS using an MPC approach, considering scenarios that represent different stages of the system's lifetime.
The NL model outperforms the LP model by providing more accurate predictions of energy losses and system capabilities,
enabling it to optimize its operational strategy to improve roundtrip efficiency and revenue.
These advantages are most significant in systems featuring batteries with high internal resistance,
like second-life systems.
However, to fully leverage the model's benefits, it is essential to identify the correct parameters.

Based on the findings of this study,
there are several opportunities for future research.
For instance, this study assumes perfect knowledge of the system state at the start of each optimization horizon, but in reality, state estimation methods are needed to determine this information accurately.
Future work could explore how the quality of state estimation affects the system optimization performance.

\bibliographystyle{ieeetr}
\bibliography{manuscript}

\end{document}